\documentclass{article}
\usepackage[T1]{fontenc}
\usepackage[utf8]{inputenc}
\usepackage{ismir,amsmath,cite,url}
\usepackage{graphicx}
\usepackage{color}
\usepackage{multirow}
\usepackage{amssymb}
\usepackage{pifont}
\usepackage{acronym}
\usepackage{comment}
\newcommand{\cmark}{\ding{51}}%
\newcommand{\xmark}{\ding{55}}%

\usepackage{lineno}

\title{Classical Guitar Duet Separation using GuitarDuets - a Dataset of Real and Synthesized Guitar Recordings}

\multauthor
{Marios Glytsos$^{1,3}$ \hspace{0.2cm} Christos Garoufis$^{1,2,3}$ \hspace{0.2cm} Athanasia Zlatintsi$^{1,2,3}$ \hspace{0.2cm} Petros Maragos$^{1,3}$} { %\bfseries{Petros Maragos$^{1,2}$ }\\
$^1$Robotics Institute, Athena Research Center, Athens, Greece \\
$^2$Institute of Language and Speech Proc., Athena Research Center, Athens, Greece \\
$^3$School of ECE, National Technical University of Athens, Athens, Greece\\
{\tt\footnotesize mariosgly@gmail.com, \{christos.garoufis,athanasia.zlatintsi\}@athenarc.gr, maragos@cs.ntua.gr}
}

\def\authorname{M. Glytsos, C. Garoufis, and A. Zlatintsi and P.Maragos}

\begin{document}

\maketitle

\begin{abstract}

Recent advancements in music source separation (MSS) have focused in the multi-timbral case, with existing architectures tailored for the separation of distinct instruments, overlooking thus the challenge of separating instruments with similar timbral characteristics. Addressing this gap, our work focuses on monotimbral MSS, specifically within the context of classical guitar duets. To this end, we introduce the GuitarDuets dataset, featuring a combined total of approximately three hours of real and synthesized classical guitar duet recordings, as well as note-level annotations of the synthesized duets. We perform an extensive cross-dataset evaluation by adapting Demucs, a state-of-the-art MSS architecture, to monotimbral source separation. Furthermore, we develop a joint permutation-invariant transcription and separation framework, to exploit note event predictions as auxiliary information. Our results indicate that utilizing both the real and synthesized subsets of GuitarDuets leads to improved separation performance in an independently recorded test set compared to utilizing solely one subset. We also find that while the availability of ground-truth note labels greatly helps the performance of the separation network, the predicted note estimates result only in marginal improvement. Finally, we discuss the behavior of commonly utilized metrics, such as SDR and SI-SDR, in the context of monotimbral MSS.

\end{abstract}

\section{Introduction}\label{sec:introduction}
\textcolor{black}{The task of music source separation (MSS) involves dissecting a musical composition into its constituent sources, typically segregating individual instruments or vocal tracks from a composite audio mixture \cite{applications_mss, music_demixing_challenge,8588410}. Due to the multitude of the co-playing sources, as well as its utility in a variety of applications~\cite{applications_mss}, MSS stands as a significant challenge in the field of Music Information Retrieval (MIR)~\cite{rafii2018overview}.} The majority of research efforts have focused on multi-timbral music source separation \cite{ht_demucs, stoter2019open, rafii2017musdb18}. In this case, the goal is the separation of distinct instrumental sources from a mixture, where the sources belong to different instrument families or types such as vocals, bass, drums and others, \textcolor{black}{and as such can be framed as an extension of the task of speech denoising into the music domain~\cite{vincent2018audio}}.

Through the \textcolor{black}{advancement of digital signal processing}~\cite{nmf} and deep learning \cite{ht_demucs,demucs_waveform}, considerable progress has been made in extracting distinct instrumental tracks from complex musical compositions. Most recent, deep-learning based approaches for MSS are divided into spectrogram-domain approaches \cite{band_split_rnn}, waveform-domain methodologies \cite{demucs_waveform, Stoller2018WaveUNetAM, jukebox_waveform, conv-tasnet, garoufis2021htmd} and hybrid ones, working simultaneously in both domains~\cite{ht_demucs}. Spectrogram-domain methods typically isolate sources via mask prediction~\cite{unet}, waveform-domain approaches enhance source separation by applying spectrogram techniques in a learned latent space~\cite{conv-tasnet, samuel2020meta}, or directly predicting the isolated waveforms~\cite{Stoller2018WaveUNetAM}, with the additional advantage of incorporating phase information, while hybrid architectures~\cite{ht_demucs} leverage the strengths of both. Moreover, recent findings have highlighted the benefits of using static or dynamic activity labels~\cite{slizovskaia2019end,petermann2020deep,score_informed_chamber,miron, informed_choral}, as well as jointly training transcription and source separation modules \cite{Lin2021AUM,jointist}, which enhances task performance, paralleling efforts in simultaneous speech recognition and separation training~\cite{speaker_joint}. 

\begin{table*}[t]
    \centering
        \begin{tabular}{|c||c|c|c|c|c|}
            \hline Datasets & Real Data Incl. & Monotimbral &  Polyphonic & Note Annotations & Duration \\
        \hline \hline musdb18~\cite{rafii2017musdb18} & \cmark & \xmark   & \cmark & \xmark & ca. 10h \\ \hline
                            MoisesDB~\cite{moisesdb} & \cmark & \xmark & \cmark & \xmark & ca. 14.5h \\ 
            \hline URMP~\cite{li2018} & \cmark & \xmark & \cmark & \cmark & 1h 6min  \\
            %\hline MedleyDB~\cite{bittner2014} \\ \hline
           \hline SLAKH~\cite{manilow2019} & \xmark & \xmark & \cmark & \cmark & ca. 145h \\ \hline
            \hline EnsembleSet~\cite{sarkar2022} & \xmark & \cmark & \xmark & \cmark & 6h 9min \\
           \hline GuitarSet~\cite{Xi2018GuitarSetAD} & \cmark & \cmark & \cmark & \cmark & 3h 3min %369 mins & 183 mins & 164 mins & 600 mins \\
           \\ \hline \hline
           GuitarDuets & \cmark & \cmark  & \cmark & Partial & 2h 44min \\ \hline
        \end{tabular}
        \vspace{-0.2cm}
    \caption{Comparison of the GuitarDuets dataset with existing datasets in the literature for music source separation; we note that GuitarSet is strictly monotimbral, since it was entirely recorded using one guitar.}
    \vspace{-0.5cm}
    \label{tab:dataset-comp}
\end{table*}

\begin{table}[!t!]
    \centering
    %\resizebox{.47\textwidth}{!}
    {%
        \begin{tabular}{|c||c|c|}
            \hline   & GuitarDuets(R) & GuitarDuets(S)\\
            \hline \hline \# Tracks & 34 & 35 \\
            \hline Dur./Track (mins)  & 1.72 $\pm 1.35$ & 3.03 $\pm 2.86$ \\
            \hline Total Dur. (mins) & 58 & 106 \\
            \hline Notes/sec. & - & 7 \\
            \hline
        \end{tabular}}
            \caption{Detailed statistics of the real and synthesized subsets of the GuitarDuets dataset; note statistics are included for the synthetic subset only.}
            \vspace{-0.55cm}
    \label{tab:guitarduet_analytics}
\end{table}

However, an area that remains relatively underexplored is monotimbral music source separation. This subfield of MSS focuses on extracting audio components that belong to the same instrument family, or \textcolor{black}{different builds of the same instrument}. It can be viewed as the counterpart of the speaker separation problem~\cite{cocktail_problem} in the music domain. \textcolor{black}{While this similarity has led to the development of similar methodologies for network training~\cite{PIT}, speaker separation, especially within the context of, rarely available in MSS datasets, multi-microphone recordings~\cite{yoshioka2018multi}, can also rely on spatial cues.} 
The limited exploration in this area can largely be attributed to the historical focus on isolating the most prominent instruments in popular music, while the demand for separation of instruments with close timbral characteristics is less pronounced. Indeed, there are very few datasets suitable for training algorithms on the task of separating instrumental tracks from the same instrument family in a polyphonic context~\cite{Xi2018GuitarSetAD}, \textcolor{black}{with the majority of publicly avalaible datasets covering the case of separating mixtures of multiple singing voices~\cite{vocal_harmony, medleyvox, schramm}}.

In this paper, we attempt to bridge this gap by introducing GuitarDuets\footnote{\textcolor{black}{The dataset is available at: {https://zenodo.org/records/12802440}}}, a dataset consisting of a total of ca.~3~hours of real and synthesized guitar duet recordings, along with partial note-level annotations, \textcolor{black}{which can be leveraged as auxiliary score information}. We benchmark GuitarDuets in the tasks of i) unconditional guitar duet separation and ii) score-informed duet separation, using the hybrid Demucs~\cite{ht_demucs} as our separation model. We also examine the possibility of integrating note-level predictions into a joint transcription and separation framework.

In more detail, the main contributions of this work are:
\begin{itemize}
    \vspace{-0.1cm}
    \item Introduction of GuitarDuets, a dataset for monotimbral music source separation, featuring both real classical guitar duet recordings and  synthetic recordings generated from online transcriptions and virtual instruments. The synthetic portion includes MIDI representations for each guitar part, enriching the dataset for algorithm training and detailed analysis.
        \vspace{-0.25cm}
\item Extensive cross-dataset evaluation across various conditions, including real and generated synthetic data, as well as the existence or absence of auxiliary score information in specific Demucs branches.

\vspace{-0.15cm}
\item Development of a joint transcription and separation framework, which incorporates transcription predictions, by adapting existing architectures~\cite{ht_demucs,rse_network} to the task of monotimbral source separation with the introduction of a permutation-invariant~\cite{PIT} loss. We show that incorporation of these note-level predictions can improve the separation of real guitar duets.
  \vspace{-0.6cm}
    \item Finally, \textcolor{black}{we analyze the behavior of} commonly-utilized source separation metrics
    in the context of classical guitar duets to understand their effectiveness when applied in sources with similar timbres.
\end{itemize}

\section{Datasets}\label{sec:datasets}

\subsection{Existing Datasets}
Datasets available for music source separation or transcription are primarily divided into multitimbral and monotimbral ones, each offering instrument-specific tracks or stems, often accompanied by transcriptions. Multitimbral datasets such as musdb18~\cite{rafii2017musdb18}, URMP~\cite{li2018}, MedleyDB~\cite{bittner2014}, MoisesDB~\cite{moisesdb} and SLAKH~\cite{manilow2019} are most prominent, featuring both real and synthesized data from a broad spectrum of instruments; some extend to multimodal forms, including for instance audiovisual \textcolor{black}{elements~\cite{li2018}}. 

In contrast, monotimbral instrumental datasets, notably fewer in number, include focused collections such as GuitarSet~\cite{Xi2018GuitarSetAD} and EnsembleSet~\cite{sarkar2022}. GuitarSet provides detailed annotations for acoustic guitar recordings, consisting of pairs of comping and soloing performances, while EnsembleSet targets chamber ensembles with high-quality synthetic reproductions of classical music. Despite their utility, these monotimbral datasets face some limitations, namely: GuitarSet's structure, with distinct solo and accompaniment parts, oversimplifies the separation task due to the distinct role of each guitar. Also, the lack of timbral differences between the two guitars \textcolor{black}{prevents the networks from focusing on timbral cues for the separation task}. On the other hand, EnsembleSet's reliance on synthetic data introduces a realism gap, underscoring the need for datasets that more accurately capture the dynamics of live musical performances. Moreover, the instruments it contains are largely monophonic, which hinders its use for scenarios with polyphonic co-playing instruments. 

\subsection{The GuitarDuets Dataset}

In this section, we will describe the GuitarDuets dataset, comprising both real recordings of classical guitar duets and synthesized recordings, leveraging virtual instruments and MIDI scores.This approach aimed to provide an original and realistic set of guitar duet recordings for training and evaluating deep learning algorithms on monotimbral MSS, while simultaneously overcoming their limited duration, granting ample training data and enabling analysis between real and synthetic datasets. In total, our dataset comprises 58.6 minutes of real data and 106 minutes of synthesized data, amounting to 164.6 minutes overall. \textcolor{black}{A comparison of the GuitarDuets with the most prominent datasets for MSS in the literature is outlined in Table 1, whereas detailed statistics about both the real and synthesized subsets of GuitarDuets are given in Table 2.}

\textbf{Real Recordings:} For the recordings featuring real classical guitars, we utilized a quiet, acoustically treated room and high-quality condenser microphones (Presonus PM-2), one for each guitar. During the recording process, four different classical guitars were used, with some tracks \textcolor{black}{(16 min.)} replayed using different guitars to further enhance timbral diversity. Simultaneous play was crucial for capturing the musical interplay between the two guitarists. The whole recording process resulted in the recording of 27 guitar duets \textcolor{black}{(per-track duration: $123 \pm 82$ sec.)}, mostly from the Modern Classical and Nuevo Tango genres and from the Romantic Period. This approach, while essential for the integrity and realism of the dataset, introduced a challenge with cross-microphone sound bleeding.
This crossover of sound presented a significant concern, as it compromises the isolation of the individual guitar tracks, impacting the quality of the dataset. In addressing the issue of source bleeding in microphones, we recorded a specialized test set that is free from such leakage. This set, consisting of 7 tracks \textcolor{black}{(per-track duration: $ 39 \pm 13$ sec.)}, was created to ensure the absence of cross-feed between microphones. Each guitar track was exported as a 44,100 Hz, 16-bit WAV file in stereo format, with mixed audio files created by averaging individual guitar performances.

\textbf{Synthetic Recordings}: \textcolor{black}{Despite the inherent realism of the real recordings, their small duration could prove problematic for network training, whereas the single recording setup utilized could import biases. A commonly utilized shortcut to increase the duration of real recordings is to virtually augment them, by synthesizing additional pieces based on note-level transcriptions and virtual music instruments, which has proven effective not only in generating multitrack datasets~\cite{manilow2019, sarkar2022}, but also in tasks such as tablature generation~\cite{synthtab,pedroza}. In our case,} {``Session Guitarist - Picked Nylon''}\footnote{https://www.native-instruments.com/en/products/komplete/guitar/ \\ session-guitarist-picked-nylon/}, a sample-based virtual instrument, was utilized to generate classical guitar sounds. It offers a wide range of playing styles, capturing the nuances of nylon-stringed guitars. We selected guitar duet MIDI scores from the {MuseScore community}\footnote{https://musescore.com/}, representing a broad spectrum of pieces. {Logic Pro X}\footnote{https://www.apple.com/logic-pro/} served as the digital audio workstation (DAW) for transforming MIDI scores into realistic guitar performances. By configuring multiple instances of the PICKED NYLON plugin with distinct timbral settings, we produced different guitar sounds. The final dataset was exported as 44,100 Hz, 16-bit WAV files in both stereo and mono formats for mixed audio file creation. 

\section{Methodology}

\subsection{Separation Architecture}
In this work the Hybrid Transformer Demucs~\cite{ht_demucs} was used \textcolor{black}{as the separation backbone}, consisting of dual U-Nets~\cite{unet}, operating in both time and spectrogram domains, each featuring four encoder and decoder layers. \textcolor{black}{The temporal encoder (TEncoder) downsamples the input waveform through a series of 1D convolutions, whereas the spectral encoder (ZEncoder) gradually compresses the STFT magnitude of the input by applying convolutions across the spectral axis.} The traditional convolutional layers, positioned between the encoder and decoder in previous iterations of the Demucs architecture~\cite{defossez2021hybrid} are replaced with a cross-domain Transformer Encoder, composed of interleaved self-attention and cross-attention Encoder layers, each equipped with \textcolor{black}{Layer Scale~\cite{touvron2021going}}. The attention mechanism operates with eight heads, and the hidden state size of the feedforward network is four times the dimension of the transformer. The primary decoder layer is shared, branching into both temporal and spectral domains, \textcolor{black}{with the respective decoders built symmetrically to the encoders}. The spectral output, post an inverse Short Time Fourier Transform (ISTFT), is merged with the temporal output, producing the model’s prediction. We note that in our experiments, the input length is set to 4 seconds.

\begin{figure}
 \includegraphics[alt={This figure shows how the binarized pianorolls are incorporated into the spectral and temporal encoders of the Demucs architecture. In more detail, for the temporal (upper) branch, the pianorolls are concatenated across the temporal dimension and inserted after the third convolutional block, whereas for the spectral (lower) branch, they are stacked and inserted after the second convolutional block.},width=\linewidth]{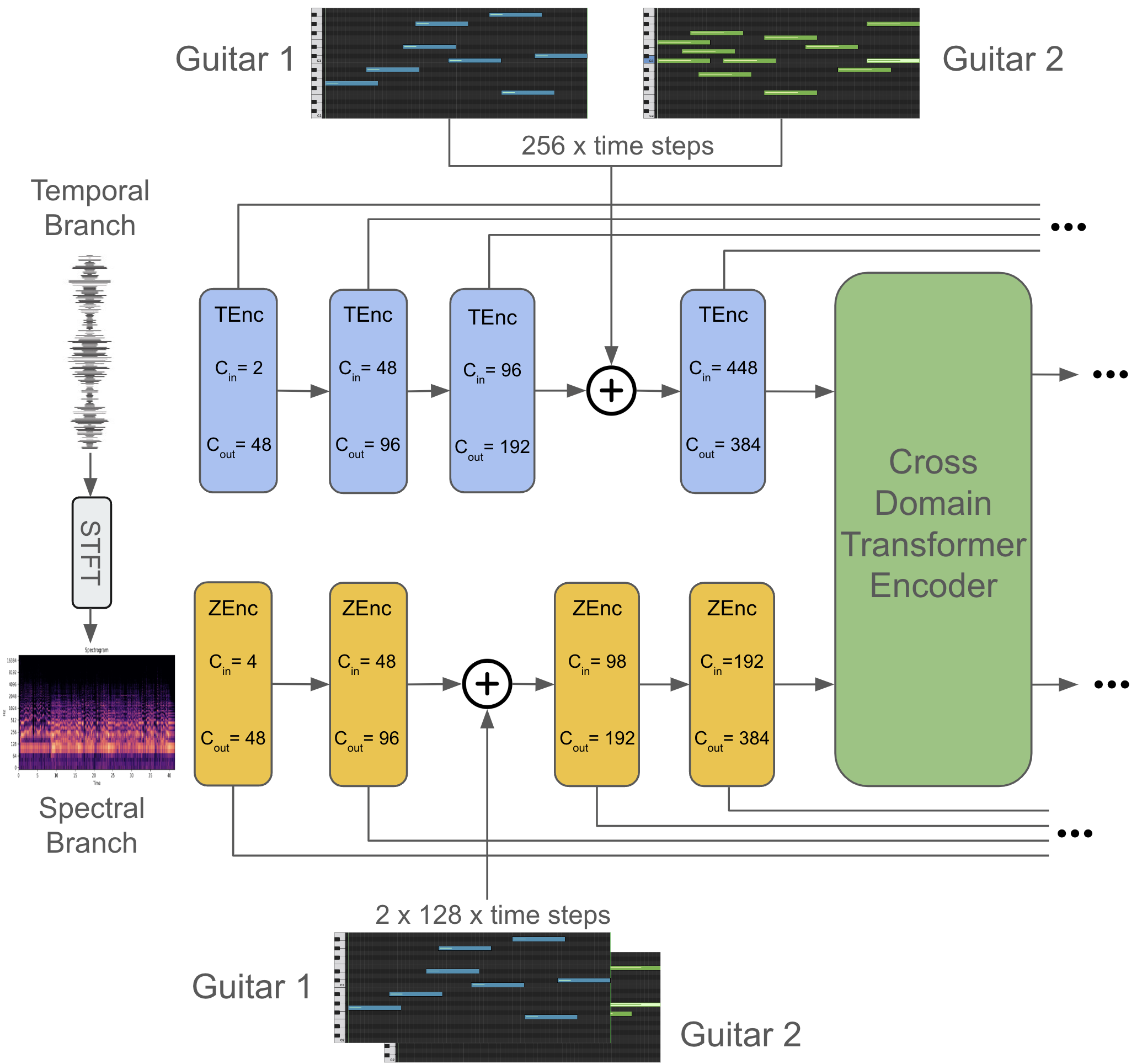}
 \vspace{-0.7cm}
 \caption{Overview of the incorporation of note-level annotations into the Demucs Architecture.}
 \vspace{-0.5cm}
 \label{fig:score_informed_demucs}
\end{figure}

\subsection{Score-Informed Separation}
In the context of Score-Informed Separation, the separation network is conditioned on the note-level transcripts of the recordings. To this end, binary vectors indicating the presence or absence of each of the 128 MIDI notes during small temporal frames are concatenated with the intermediate feature maps in each branch, as indicated in Fig.~\ref{fig:score_informed_demucs}. In particular, in the temporal branch, the activity labels of each guitar are inserted after the third TEncoder layer. The binary vector for each guitar has a dimensionality of \(128 \times N_s\), where $N_s$ corresponds to the number of \text{samples for each 4-second segment}, yielding a combined shape of \(256 \times N_s\) for both guitars. 
Thus, the activity labels have to be downsampled across the temporal axis, 
to match the resolution of the encoder at this stage. In a parallel manner, within the frequency branch, these binary vectors are introduced following the second ZEncoder layer. The shape of the activity labels for concatenation is \(2 \times 128 \times N_s\), aligning with the two guitars' MIDI notes. In this case, the activity labels are concatenated with the feature maps across the channel dimension; since the frequency resolution of the ZEncoder, at this stage, matches the number of MIDI notes, resampling occurs only across the temporal axis, by downsampling the note activity labels to the respective temporal resolution of the ZEncoder. 

\begin{figure}%[t]
\includegraphics[alt={This figure is illustrating the architecture of the trascription-separation pipeline. Two guitars are shown at the top, each connected to a microphone, capturing their audio. The audio signals from both guitars are combined into a single waveform. This combined waveform is input into a transcription model, which processes the audio and produces separate transcriptions for Guitar 1 and Guitar 2. The transcribed data for both guitars together with the initial mixed audio signal are fed into a supervised source separation model. This model separates the original audio waveform into two distinct waveforms, one for each guitar.},width=\linewidth]{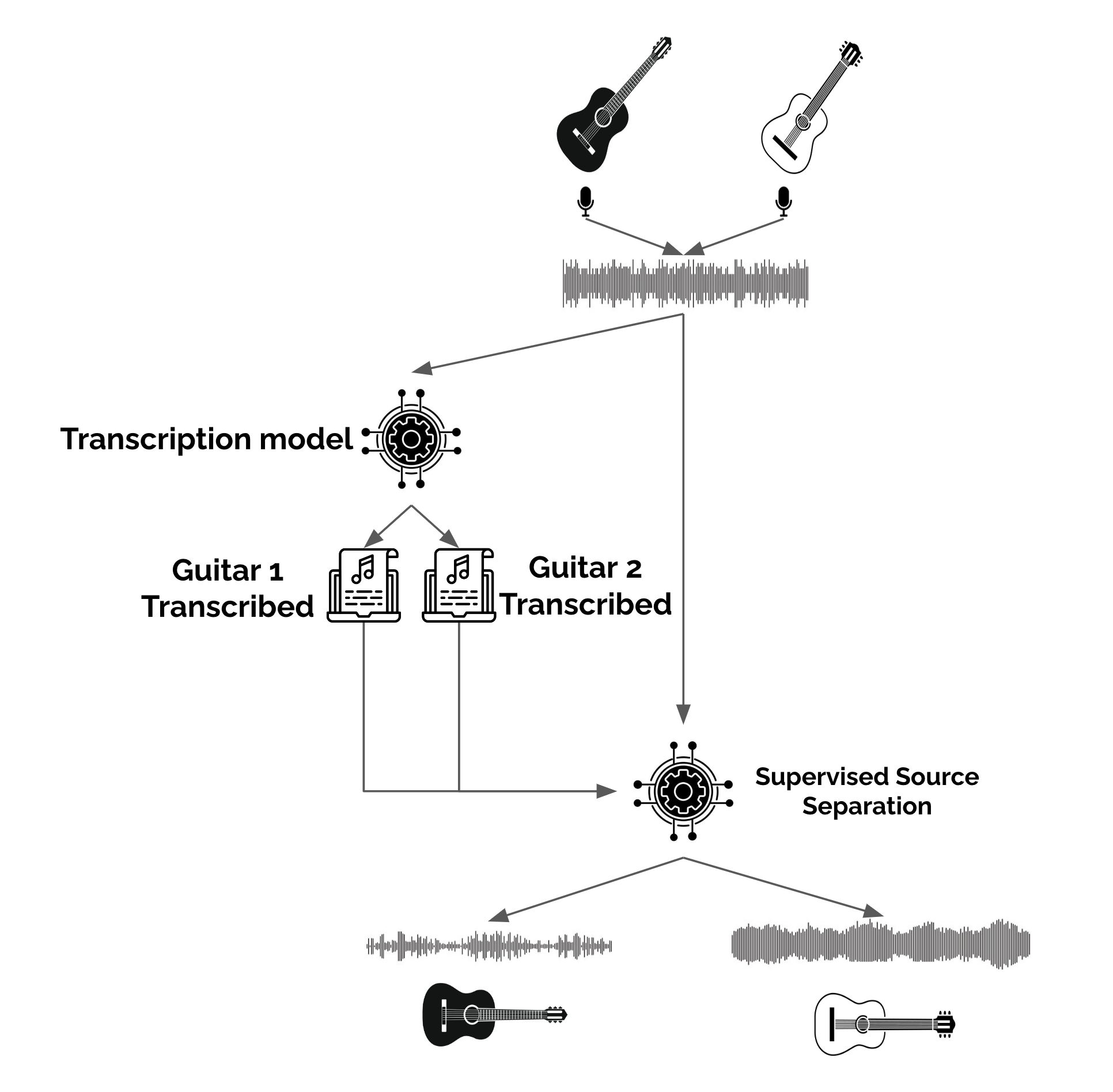}
 \caption{Overview of the proposed methodology for joint transcription and separation of guitar duets.}
 \vspace{-0.5cm}
 \label{fig:pipeline_transc_sep}
\end{figure}

If transcriptions are not available, we use a separate transcription network to generate them creating a joint transcription-separation framework. The first network would intake the combined sounds of the two guitars and generate a binarized piano roll representation for each individual guitar. Afterward, the second model combines the mixed audio and the generated piano rolls to create separate estimates

for each guitar as depicted in Figure~\ref{fig:pipeline_transc_sep}. From a musical endpoint the transcription network could potentially capture note interdependencies and guitar duet patterns through binarized vector features, aiding in note prediction. This transcription informs the separation model, which refines the output by focusing on timbre.

For the transcription architecture, we utilize the Residual Shuffle-Exchange Network (RSE) \cite{rse_network}, which has achieved state-of-the-art results in MusicNet~\cite{thickstun2016learning}. This network enhances the neural Shuffle-Exchange network \cite{neural_shuffle_exchange} by employing both Switch and Shuffle layers to capture sequence dependencies effectively, as well as reducing its computational overhead by incorporating strided convolutions. For further details about the architecture we refer to \cite{rse_network,neural_shuffle_exchange}. In our implementation, the RSE's output layer is modified to produce a binarized $2 \times 128$-dimensional representation, to assign activity labels  for each of the 128 MIDI notes to the corresponding instrument.
\section{Experimental Evaluation}

\subsection{Experimental Setup}

For the separation experiments, we used both the real and synthesized subsets of GuitarDuets, which we will further denote as GuitarDuets(R) and GuitarDuets(S), respectively, as well as the GuitarSet, for which mixtures were generated via addition of the available comping and solo excerpts. We adapted the backbone Demucs model for classical guitar duet separation, modifying it to output two stereo signals, one for each guitar. %The model inferences 4-second segments to balance computational efficiency and musical nuance capture. 
Data augmentation techniques including channel swapping, time cropping, amplitude scaling and remixing individual guitar parts from different performances %FlipChannels, Shift, Remix and Scale 
were employed \textcolor{black}{during network training} to improve generalization. %The model operates at a 44.1~kHz sample rate, maintaining original recording fidelity.
As our loss function we used the quantity:
\vspace{-0.25cm}
\begin{equation*}
    \alpha \cdot \mathrm{min}(|\hat{g_1} - g_1| + |\hat{g_2} - g_2|, |\hat{g_2} - g_1| + |\hat{g_1} - g_2|) 
\end{equation*}
%\vspace{-0.2cm}
\begin{equation}
+\beta \cdot |(\hat{g_1} + \hat{g_2}) - (g_1 + g_2)|,
\end{equation}
%\vspace{-0.2cm}
%
where the first term corresponds to the traditional permutation-invariant L1 loss between the ground truth signals $g_1$, $g_2$ and the output sources $\hat{g_1}$, $\hat{g_2}$, and the second term models the similarity between the sum of the guitar estimates and the input mixtures, \textcolor{black}{encouraging the network to provide separate guitar tracks which neither discard nor duplicate note instances}, whereas the weight values were set, \textcolor{black}{after preliminary experiments}, to $\alpha = 0.8$, $\beta = 0.2$. 

For the transcription architecture experiments, we employed GuitarSet and the GuitarDuets(S) dataset, which contain note-level annotations for individual guitar parts. We transformed labels from GuitarSet (.jams files) and the MIDI files from our dataset to CSV format. All audio files, initially sampled at 44,100 Hz, were resampled to 11,000 Hz and converted to mono, to render them compatible with the RSE backbone~\cite{rse_network}. Similar to the separation case, the loss function --in this case, the binary cross entropy-- was employed within a permutation invariant framework.

\begin{table*}[t]
    \centering
    \makebox[\textwidth]{%
    \begin{tabular}{|c|c|c||c|c|c|c|}
        \hline
    \multicolumn{3}{|c||}{Source Datasets} & \multicolumn{4}{c|}{Metrics} \\ \hline GuitarDuets(R) & GuitarDuets(S) & GuitarSet & SDR & SI-SDR & SAR & SIR \\
        \hline \hline
        \multirow{2}{*}{\cmark} & \multirow{2}{*}{\cmark} & \multirow{2}{*}{\cmark} & G1: 4.297 & G1: 3.403 & G1: 7.670 & G1: 10.766 \\
                    & &               & G2: 0.835 & G2: -2.880 & G2: 2.062 & G2: 4.495  \\
        \hline
                \multirow{2}{*}{\cmark} & \multirow{2}{*}{\xmark} & \multirow{2}{*}{\cmark} & G1: 4.522 & G1: 4.280 & \textbf{G1: 9.483} & G1: 6.273  \\
                                & & & G2: 1.359 & G2: -2.238 & \textbf{G2: 10.898} & G2: 7.631\\
        \hline
                \multirow{2}{*}{\xmark} & \multirow{2}{*}{\cmark} & \multirow{2}{*}{\cmark} & G1: 4.493 & G1: 1.530 & G1: 8.191 & G1: 7.038  \\
                                & & & G2: 1.137 & G2: -1.566 & G2: 8.305 & G2: 8.081\\
        \hline
                \multirow{2}{*}{\xmark} & \multirow{2}{*}{\xmark} & \multirow{2}{*}{\cmark} & G1: 4.632 & G1: 3.871 & G1: 7.971 & G1: 7.321 \\
                            & & & \textbf{G2: 1.378} & \textbf{G2: -1.198} & G2: 6.332 & \textbf{G2: 8.968} \\
        \hline
                \multirow{2}{*}{\xmark} & \multirow{2}{*}{\cmark} & \multirow{2}{*}{\xmark} & G1: 3.472 & G1: 1.857 & G1: 8.212 & G1: 8.217 \\
                             & & & G2: 0.200 & G2: -4.052 & G2: 4.502 & G2: 4.501  \\
        \hline
                \multirow{2}{*}{\cmark} & \multirow{2}{*}{\xmark} & \multirow{2}{*}{\xmark} & G1: 4.952 & G1: 3.573 & G1: 7.628 & G1: 10.413  \\
                             & & & G2: 1.014 & G2: -3.536 & G2: 1.424 & G2: 4.873  \\
        \hline
                \multirow{2}{*}{\cmark} & \multirow{2}{*}{\cmark} & \multirow{2}{*}{\xmark}  & \textbf{G1: 5.882} & \textbf{G1: 4.315} & G1: 8.488 & \textbf{G1: 11.706}  \\
                         & &       & G2: 0.920 & G2: -3.133 & G2: 0.896 & G2: 4.104  \\
        \hline
    \end{tabular}}
    \vspace{-0.3cm}
        \caption{Separation results on the testing set of GuitarDuets(R), according to the datasets utilized during training. G1 corresponds to the 1st guitar, G2 to the 2nd. Higher is better for all metrics.}
        \vspace{-0.3cm}
\end{table*} 

\begin{table*}[t]
    \centering
    \begin{tabular}{|c|c|c|c|c|c|c|c|}
    \hline
    \multirow{2}{*}{Dataset} & \multirow{2}{*}{Note Labels}& \multicolumn{2}{c|}{Branch Conditioning} & \multicolumn{4}{c|}{Metrics} \\
    \cline{3-8} & & Time & Frequency & SDR & SI-SDR & SAR & SIR \\ \hline \hline
    \multirow{6}{*}{GuitarDuets(S)} &
    \multirow{6}{*}{Ground Truth} & \multirow{2}{*}{\cmark} & \multirow{2}{*}{\xmark} & G1: 4.453 & G1: 3.117 & G1: 4.972 & \textbf{G1: 12.411}   \\
                        & & & & G2: 4.355 & G2: 0.072 & G2: 3.197 & G2: 8.292 \\ \cline{3-8}
                        & & \multirow{2}{*}{\xmark} & \multirow{2}{*}{\cmark} & G1: 4.547 & G1: 3.293 & \textbf{G1: 4.685} & G1: 9.523   \\
                        & & & & G2: 3.301 & G2: -0.410 & G2: 3.451 & G2: 9.882 \\ \cline{3-8}
                        & & \multirow{2}{*}{\cmark} & \multirow{2}{*}{\cmark} & \textbf{G1: 4.717} & \textbf{G1: 3.378} & G1: 4.362 & G1: 12.081  \\
                        & & & & \textbf{G2: 4.863} & \textbf{G2: 0.154} & \textbf{G2: 4.316} & \textbf{G2: 10.537}  \\ \hline \hline
        \multirow{4}{*}{GuitarDuets(S)}& \multirow{2}{*}{Estimated} & \multirow{2}{*}{\cmark} & \multirow{2}{*}{\cmark} & \textbf{G1: 3.414} & G1: 1.398 & G1: 3.455 & G1: 10.655 \\
                        & & & & G2: 1.977 & \textbf{G2: -1.511} & G2: 3.087 & \textbf{G2: 7.035} \\ \cline{2-8} &
    \multirow{2}{*}{None} & \multirow{2}{*}{\xmark} & \multirow{2}{*}{\xmark} & G1: 2.575 & \textbf{G1: 2.436} & \textbf{G1: 4.473} & \textbf{G1: 12.795}  \\
                        & &  & & \textbf{G2: 2.569} & G2: -2.514 & \textbf{G2: 3.473} & G2: 5.717  \\ \hline \hline
                            \multirow{4}{*}{GuitarDuets(R)} & \multirow{2}{*}{Estimated} & \multirow{2}{*}{\cmark} & \multirow{2}{*}{\cmark} & \textbf{G1: 5.313} & \textbf{G1: 4.352} & \textbf{G1: 7.638} & \textbf{G1: 11.110} \\
                        & & & & \textbf{G2: 1.035} & \textbf{G2: -3.291} & \textbf{G2: 1.998} & \textbf{G2: 5.089} \\ \cline{2-8}
    & \multirow{2}{*}{None} & \multirow{2}{*}{\xmark} & \multirow{2}{*}{\xmark} &  G1: 4.952 & G1: 3.573 & G1: 7.628 & G1: 10.413   \\
                        & & & & G2: 1.014 & G2: -3.536 & G2: 1.424 & G2: 4.873 \\ \hline
    \end{tabular}
    \vspace{-0.3cm}
    \caption{Separation results on the testing sets of GuitarDuets(S), GuitarDuets(R), when using solely the respective training sets for training, depending on the availability of note-level annotations and the Demucs branches conditioned on them. G1
corresponds to the 1st guitar, G2 to the 2nd. Higher is better for all metrics.}
    \label{tb:labels_ablation}
    \vspace{-0.5cm}
\end{table*}

\subsection{Cross-Dataset Analysis}

For the purposes of the cross-dataset analysis, we consider the GuitarDuets(R) and GuitarDuets(S) subsets as separate datasets, and train the Demucs backbone on various combinations of GuitarSet, GuitarDuets(R) and GuitarDuets(S), using the same experimental protocol \textcolor{black}{and an 80-20 training-validation split}; all networks are evaluated on the \textcolor{black}{bleeding-free} testing set of GuitarDuets(R). Upon inspection of the results, presented in Table~3, several key insights emerge. Namely, the complete GuitarDuets dataset yielded the highest SDR values for the first guitar. 
The inclusion of the synthesized subset likely provided additional information that enhanced the model’s performance with regards to the SDR. On the other hand, the inclusion of these synthetic parts made the model prone to auditory artifacts, since the highest SAR scores were achieved for the combination of GuitarDuets(R) with the GuitarSet. Finally, we observe that the combination of the complete GuitarDuets dataset with GuitarSet leads in diminished performance, probably due to the structural differences between the training subsets.

In our analysis, we observed a consistent discrepancy in the Signal-to-Distortion Ratio (SDR) between the two guitars, where the first guitar exhibited a decent SDR, while the second often fell below a threshold of 1 dB. This pattern suggests that the algorithm may be effectively separating the first guitar by identifying it as the primary source, whereas it perceives the second guitar as background noise, or merely an auditory artifact. \textcolor{black}{It is important to note that the average amplitude of both guitars is on the same scale, so this observation is not attributed to amplitude differences.} Notably, we observe that the most consistent SDR values for G2 were achieved when GuitarSet was included in the training set, which we attribute to its relatively noise-free structure.

\subsection{Score-Informed Separation Approaches}

For the experiments concerning score-informed separation, we investigated the integration of activity labels into our network, considering Demucs' operation across frequency and temporal domains, by using the GuitarDuets(S) as our dataset since it contains note-level annotations. We also investigate, using both GuitarDuets(R) and GuitarDuets(S), whether the joint transcription-separation architecture can aid in effective separation in scenarios where no ground truth data is available. \textcolor{black}{In both cases, a part of the dataset (the bleeding-free subset of GuitarDuets(R), and 10\% of GuitarDuets(S)) was used for performance evaluation; the rest were used for training and validation, at an 80:20 ratio}.

The analysis, as detailed in Table~\ref{tb:labels_ablation}, reveals that while using the temporal branch for note label integration leads to slightly improved results compared to the spectral branch, the hybrid approach achieves the most promising outcomes. This performance can be attributed to the inherent design of the Demucs architecture, which has historically shown improved efficiency when leveraging both domains concurrently~\cite{defossez2021hybrid}. It is also noteworthy that while the integration of ground-truth labels leads to higher SIR values, presumably due to the guidance that these labels provide to the separation network about the identity of each guitar, the improvement in SAR is marginal.

\begin{figure}[t]
\centering
 \includegraphics[alt={This figure shows (on top) a spectrogram of an isolated guitar excerpt with the score-informed Demucs,},width=0.97\columnwidth]{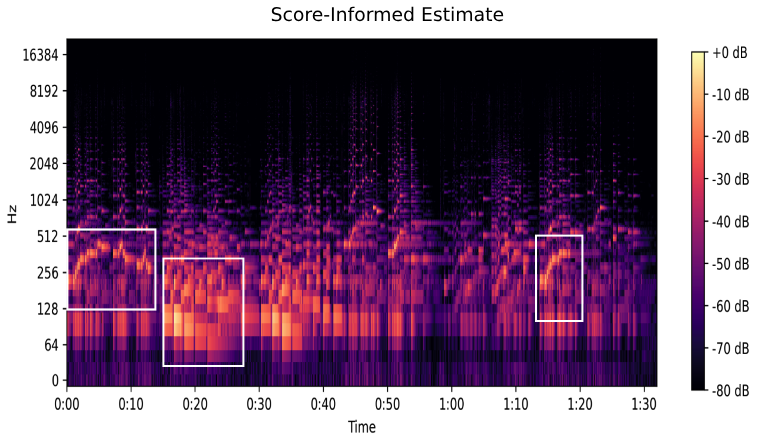}
 \includegraphics[alt={while (at bottom) a spectrogram of the same excerpt isolated using the variant of DeMucs that is conditioned on note predictions. The rectangles indicate excerpts of the signal where the score-informed variant performs better, due to its ability to better sustain individual notes.},width=0.97\columnwidth]{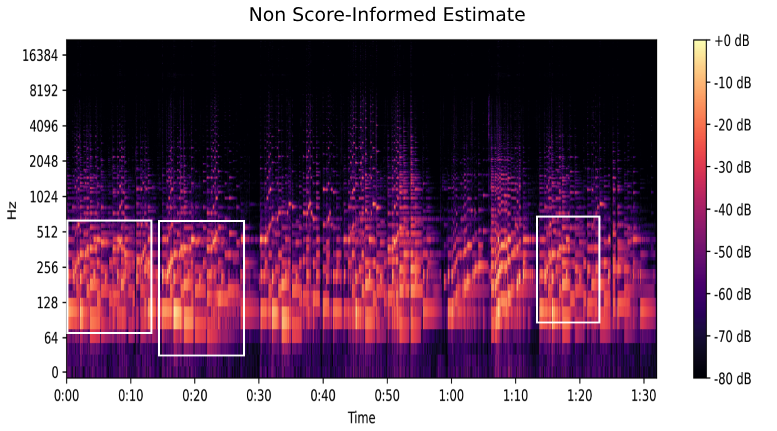}\\
 \vspace{-0.45cm}
 \caption{Comparison of spectrogram estimates with estimated (top) note-level annotations, and without those (bottom), for an instance from GuitarDuets (R).}
 \vspace{-0.65cm}
 \label{fig:score_informed_ex}
\end{figure}

Regarding the joint transcription and separation framework, the performance does not reach the levels achieved when ground-truth note-level annotations are available, as measured in GuitarDuets(S). On the other hand, while in the case of GuitarDuets(R), the performance is slightly improved when these pseudo-annotations are available, GuitarDuets(S) achieves better results in their absence. A comparison of guitar estimates for the models trained with and without predicted note label information, for an instance of the GuitarDuets(R) test set, can be depicted in Figure~\ref{fig:score_informed_ex}; we assume that the incorporation of note activity labels enables the separation model to more accurately sustain notes, enhancing the quality of the isolated melodic and accompanying parts. On the other hand, we attribute the performance drop, when using GuitarDuets(S), to the reduced generalization of the transcription network. Since the training set of GuitarDuets(S) was used for its training, the separation network was trained using mostly correct labels, but evaluated with note-level annotations of pieces the transcription network did not use for training. 

\subsection{Comparative Metric Analysis} \label{subsec:analysis}

In the field of MSS, the evaluation of separation quality is often quantified using metrics such as SDR~\cite{vincent2006performance} and SI-SDR~\cite{le2019sdr}. While they have been extensively used in studies focusing on separating different instruments, their behavior on sources with similar timbral characteristics remains less explored. Given that most prior work involves instruments with distinct timbres, direct comparison of our results with SDR values achieved across different datasets may not be appropriate for our study, which focuses on two classical guitars with similar timbral properties. 

In order to identify potential disparities in the behavior of the metrics that can be attributed to timbral similarities in the mixture components, we simulated imperfect estimates of a reference signal $x_1$ by creating additive synthetic mixtures of the signals $x_1$, $x_2$ as:
\begin{equation}
    m = \alpha x_1 + (1-\alpha)x_2,
\end{equation}
with $\alpha \in (0,1)$, and measured the values of the SDR, SI-SDR metrics between these mixtures and $x_1$. We examined two cases using signals derived from Track 29 of the GuitarDuets(S): i) a monotimbral mixture, where both $x_1$ and $x_2$ constitute guitar signals, and ii) a multitimbral mixture, where $x_2$ was synthesized from the second guitar's notes using a piano virtual instrument plugin. To guarantee a fair comparison across all tests, we performed amplitude normalization between the two tracks for each experiment.

\begin{figure}
    \centering
    \includegraphics[alt={In the left figure, the Signal-to-Distortion Ratios between a guitar-guitar and a guitar-piano mixture are compared, at various mixing ratios between the two instruments. The figure indicates that for a given mixing ratio, the guitar-guitar mixture achieves a higher SDR than the guitar-piano one. The right figure displays the same comparison with the SI-SDR, through which similar insights can be derived},width=0.98\columnwidth]{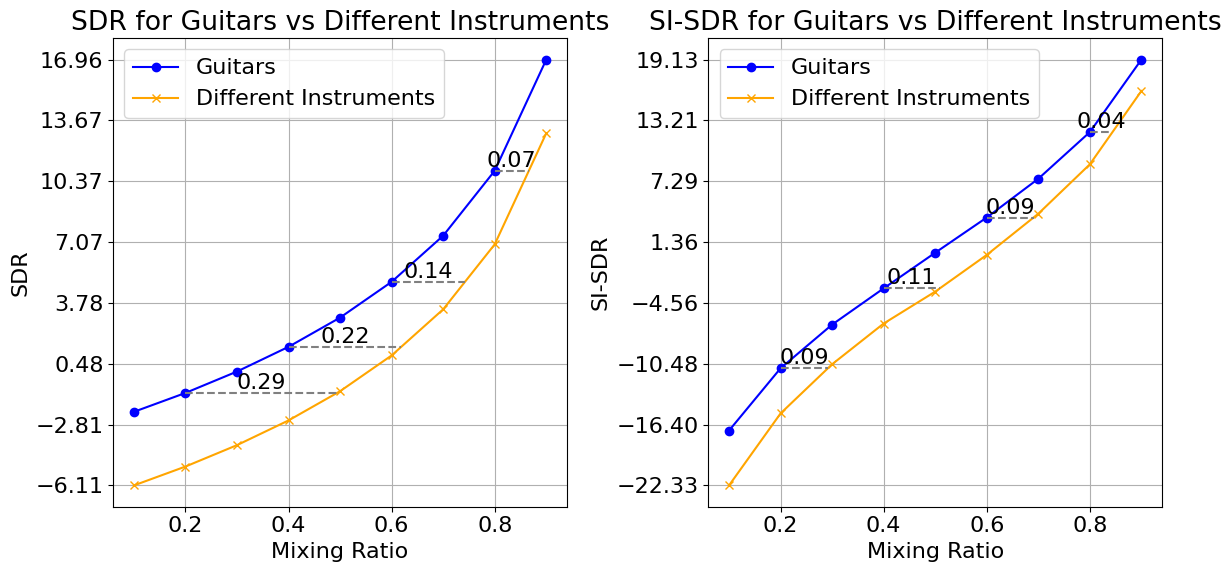}
    \vspace{-0.35cm}
    \caption{Comparison of the behavior of SDR (left) and SI-SDR (right) when assessing the separation of monotimbral (blue line) or multitimbral (orange line) duets.}
    \label{fig:comparison_results}
    \vspace{-0.5cm}
\end{figure}

The results, displayed in Figure \ref{fig:comparison_results}, indicate that both metrics for the guitar mixtures are consistently higher than those obtained from mixtures of different instruments. For instance, the mixing ratio $\alpha$ required to reach an SDR value of 5 approaches 0.8 for the multi-timbral case, while 0.6 for the mono-timbral case. This suggests that the timbral similarity between the two guitars introduces a challenge for the metrics to accurately assess the quality of separation.

\section{Conclusions}

In this paper, we introduced GuitarDuets, a dataset consisting of both real and synthesized classical guitar duets. We exhibit that our dataset can be utilized for developing monotimbral source separation algorithms within both traditional and score-informed frameworks. We further developed a joint permutation-invariant framework for transcription and separation of monotimbral mixtures, which we show that can lead to improved performance in separation of real guitar duets. In the future, we plan to extend the recordings of both the real and synthesized subsets of GuitarDuets, and provide note-level annotations for its real subset. \textcolor{black}{Furthermore, regarding the joint transcription-separation architecture, we intend to explore more sophisticated ways for integrating the predicted guitar transcripts into the separator~\cite{perez2018film, meseguer2019conditioned}}. Finally, we aim to conduct extensive listening tests, which will help in further shedding light into both the performance of the various approaches we compare, and the significance of objective metrics within the context of monotimbral audio source separation.  

\textbf{Acknowledgments:} This research was supported by the Hellenic Foundation for Research and Innovation (H.F.R.I.) under the ``3rd Call for H.F.R.I. Research Projects to support Post-Doctoral Researchers'' (Project Acronym: i-MRePlay, Project Number: 7773).

\bibliography{ISMIRtemplate}

\end{document}